\documentstyle[pre,aps,twocolumn,epsf]{revtex}
\begin{document}
\title{Conformal Dynamics of Fractal Growth Patterns Without Randomness}
\author{Benny Davidovitch,$^1$ M.J. Feigenbaum, $^2$, H.G.E Hentschel,$^3$ and Itamar
Procaccia,$^1$}
\address{$^1$ Dept. of Chemical Physics, The Weizmann Institute of
Science, Rehovot 76100, Israel\\
$^2$ The Rockefeller University, 1230 York Ave. New York, NY
10021\\
$^3$ Department of Physics, Emory University, Atlanta, GA, 30322}

\date{version of \today}
\maketitle

\begin{abstract}
Many models of fractal growth patterns (like Diffusion Limited Aggregation
and Dielectric Breakdown Models) combine complex
geometry with randomness; this double difficulty is a stumbling block
to their elucidation. In this paper we introduce a wide class of
fractal growth models with highly complex geometry but without any
randomness in their growth rules. The models are defined in terms
of deterministic itineraries of
iterated conformal maps, automatically generating the function $\Phi^{(n)}(\omega)$
which maps the exterior of the unit circle to the exterior of an $n$-particle
growing aggregate. The complexity of the evolving interfaces is fully contained 
in the deterministic dynamics of the conformal map $\Phi^{(n)}(\omega)$. We
focus attention to a class of growth models in which the itinerary is quasiperiodic.
Such itineraries can be approached via a series of rational approximants. The
analytic power gained is used to introduce a scaling theory of the fractal growth patterns.
We explain the mechanism for the fractality of the clusters and identify the 
exponent that determines the fractal dimension. 
\end{abstract}

\pacs{PACS numbers: ...}
\section{Introduction}
In this paper we introduce a new class of fractal growth patterns 
in two dimensions, constructed in terms of the conformal maps from the exterior
of the unit circle to the exterior of the growing cluster. Until now
most of the interesting fractal growth models included
randomness as an essential aspect of the growth algorithms. 
Foremost in such models has been the diffusion limited aggregation (DLA) model that
was introduced in 1981 by T. Witten and L. Sander \cite{81WS}. This model
has been shown to underlie many pattern forming processes including
dielectric breakdown \cite{84NPW}, two-fluid flow \cite{84Pat}, and
electro-chemical deposition \cite{89BMT}. The algorithm begins with fixing
one particle at the center of coordinates in $d$-dimensions, and follows
the creation of a cluster by releasing random walkers from infinity,
allowing them to walk around until they hit any particle belonging to the
cluster. Upon hitting they are attached to the growing cluster. The
growth probability for a random walker to hit the interface is known as
the ``harmonic measure", being the solution of the harmonic (Laplace) equation
with the appropriate boundary conditions. 
The DLA model was generalized to a family of models known collectively as
Dielectric Breakdown Models, in which the density of growth probability is the
density of the harmonic measure raised to a power $\eta$ \cite{95Piet}. 
For $\eta=1$ one regains the
DLA model; the interval $\eta\in (0,\infty)$ generates a family of growth
patterns from compact to a single needle. For $\eta=0$ one obtains a
growth probability that is uniform for all boundary points. This is known
as the Eden model that was introduced originally to describe the growth
of cancer cells \cite{88Mea}.

The fundamental difficulty of all these models is that their mathematical
description calls for solving equations with boundary conditions on a
complex, evolving interface. It is therefore advantageous to swap for a
simple boundary, like the unit circle, and to delegate the complexity to
the dynamics of the conformal map from the exterior of the unit circle to
the exterior of the growing cluster. For continuous time processes this
method had been around for decades \cite{58ST,84SB}, and had been used extensively. For
discrete particle growth such a language was developed recently 
\cite{98HL,99DHOPSS,99DP}, 
showing that DLA in two dimensions can be grown by iterating stochastic conformal
maps. In this paper we employ this new language to define models in which
the stochasticity is eliminated altogether, to create deterministic iterations of
conformal maps with very interesting fractal growth properties. It is stressed
below that these new models and their interesting properties are natural
extensions of the discrete conformal dynamics; it may be very difficult
to study such models with the traditional techniques in physical space.

A central thesis of this paper is that the growth models introduced 
below are simpler to understand than DLA, even though the fractal
geometry exhibited does not seem simpler. Indeed, we present below
some tools and concepts that allow us to explain why the growing
cluster is fractal. We present a scaling theory of the growing
clusters, and identify the exponent that determines the fractal
dimension. 
In sect. 2 we review the basic ideas of conformal dynamics as
a method to grow DLA and related growth patterns. In
Sec.3 we make the point that within this framework randomness can
be eliminated from the discussion without changing the properties
of the fractal growth: one can have deterministic growth rules
with clusters that are indistinguishable from DLA. In Sec.4 we introduce
fractal growth patterns that are obtained from quasiperiodic itineraries
of iterated conformal maps. These itineraries are characterized by
a winding number $W$. The growing clusters have complex geometries and
a difference appearance for every $W$. We propose nevertheless that
all the quadratic irrationals belong to the same universality class,
and that the dimensions of their clusters are the same. In Sec.5 we 
consider rational approximants $P/Q$ to the quadratic irrational 
winding numbers $W$'s. With 
rational approximants the growth patterns crossover from a fractal
phase of growth to a 1-dimensional star-like growth pattern. We argue
that the analysis of the crossover as a function of $Q$ provides us
with a scaling theory, allowing the introduction of universality classes
and the achievement of data collapse. In Sec.6 we elucidate the mechanism for
crossover from fractal to 1-dimensional growth, and identify the
exponent that determines the fractal dimension. In Sec.7 we summarize and offer final
remarks regarding the availability of a renormalization group treatment
and of the road ahead.
\section{Discrete Conformal Dynamics for Fractal Growth Patterns}

The basic idea is to follow the evolution of the conformal mapping
$\Phi^{(n)}(w)$ which maps the exterior of the unit circle $e^{i\theta}$ in the
mathematical $w$--plane onto the complement of the (simply-connected)
cluster of $n$ particles in the physical $z$--plane \cite{98HL,99DHOPSS,99DP}. 
The unit circle is
mapped to the boundary of the cluster which is parametrized by the arc
length $s$, $z(s)=\Phi^{(n)}(e^{i\theta})$. This map $\Phi^{(n)}(w)$ is
made from compositions of elementary maps $\phi_{\lambda,\theta}$,
\begin{equation}
\Phi^{(n)}(w) = \Phi^{(n-1)}(\phi_{\lambda_{n},\theta_{n}}(w)) \ ,
\label{recurs}
\end{equation}
where the elementary map $\phi_{\lambda,\theta}$ transforms the unit 
circle to a circle with a ``bump" of linear size $\sqrt{\lambda}$ around
the point $w=e^{i\theta}$. Accordingly the map $\Phi^{(n)}(w)$ adds on a
new bump to the image of the unit circle under $\Phi^{(n-1)}(w)$. The
bumps in the $z$-plane simulate the accreted particles in
the physical space formulation of the growth process. The main idea 
in this construction is to
choose the positions of the bumps $\theta_n$ and their sizes $\sqrt{\lambda_n}$
such as to achieve accretion of {\em fixed linear size} bumps on the
boundary of the growing cluster according to the growth rules appropriate
for the particular growth model that we discuss.

\begin{figure}
\hskip -1.5 cm
\epsfxsize=7.0truecm
\epsfysize=7.0truecm
\epsfbox{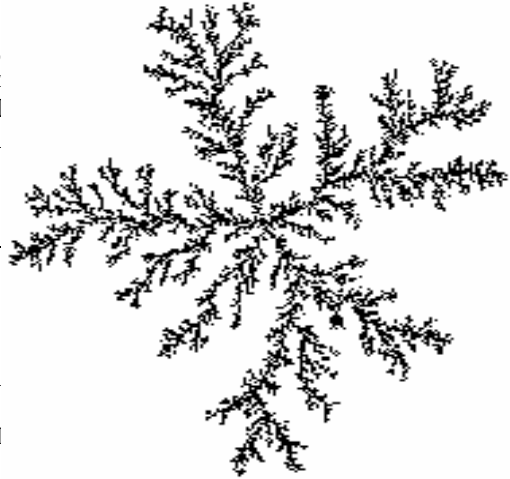}
\caption{A DLA cluster, $n=10^5$.}
\label{DLAcluster}
\end{figure}

As an example consider DLA. In $z$-space we want to accrete particles 
according to the harmonic measure. This means that the probability for the
$n$th particle to hit a boundary element
$ds$ equals $P(s) ds$, where $P(s)$ (the density of the harmonic measure 
\cite{98HL,99DHOPSS,86HMP}) and $ds$ are:
\begin{eqnarray}
P(s)&=&\frac{1}{|{\Phi^{(n-1)}}^{'}(e^{i\theta})|} \ ,\\
ds&=&|{\Phi^{(n-1)}}^{'}(e^{i\theta})| d\theta \ .
\end{eqnarray} 
Here $e^{i\theta}$ is the preimage of $z(s)$. Accordingly the probability to grow
on an interval $d\theta$ is uniform (independent of $\theta$). Thus to grow a DLA we
have to choose random positions $\theta_n$, and
$\lambda_n$ in Eq.(\ref{recurs}) according to
\begin{equation}
\lambda_{n} = \frac{\lambda_0}{|{\Phi^{(n-1)}}' (e^{i \theta_n})|^2} \ .
\label{lambdan}
\end{equation}
This way we accrete fixed size bumps in the physical plane according
to the harmonic measure. The elementary map $\phi_{\lambda,\theta}$ is chosen as \cite{98HL}
\begin{eqnarray}
&&\phi_{\lambda,0}(w) = w^{1-a} \left\{ \frac{(1+
\lambda)}{2w}(1+w)\right. \nonumber\\
&&\left.\times \left [ 1+w+w \left( 1+\frac{1}{w^2} -\frac{2}{w} \frac{1-
\lambda} {1+ \lambda} \right) ^{1/2} \right] -1 \right \} ^a \\
&&\phi_{\lambda,\theta} (w) = e^{i \theta} \phi_{\lambda,0}(e^{-i \theta}
w) \,,
\label{eq-f} 
\end{eqnarray}
The parameter $a$ is confined in the range $0< a < 1$, determining
 the shape of the bump. We employ $a=2/3$ which is consistent with
semicircular bumps. The recursive dynamics can be represented as iterations 
of the map $\phi_{\lambda_{n},\theta_{n}}(w)$,
\begin{equation}
\Phi^{(n)}(w) =
\phi_{\lambda_1,\theta_{1}}\circ\phi_{\lambda_2,\theta_{2}}\circ\dots\circ
\phi_{\lambda_n,\theta_{n}}(\omega)\ . \label{comp}
\end{equation}
The DLA cluster is fully determined by the stochastic itinerary
$\{\theta_i\}_{i=1}^n$. In Fig.~\ref{DLAcluster} we present a typical DLA cluster
grown by this method to size $n=10^5$. 
The main point of this paper is that the same method can be now used to
grow a large variety of interesting fractal shapes, but without any
randomness in the growth algorithm.
\section{DLA-like clusters without randomness}

\begin{figure}
\hskip -1.5 cm
\epsfxsize=7.0truecm
\epsfysize=7.0truecm
\epsfbox{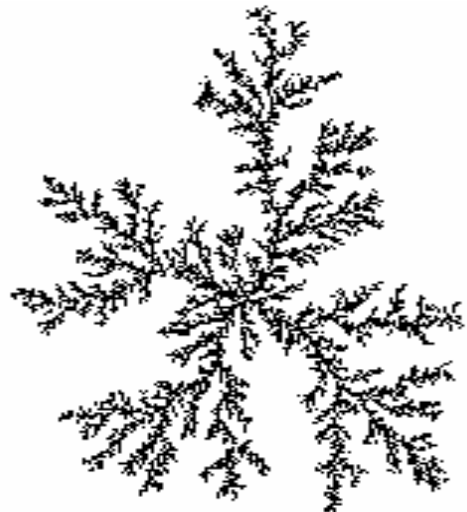}
\caption{A Bernoulli cluster, $n=10^5$.}
\label{Bercluster}
\end{figure}

As a first example of a new model we will remove the stochasticity of
DLA, leaving the growth characteristics unchanged. To this aim consider
an itinerary
\begin{equation}
\theta_{n+1}=2\theta_n ~~{\rm mod} 2\pi \ , \label{bernoulli}
\end{equation}
together with Eq.(\ref{lambdan}). Such an itinerary, although deterministic, 
is chaotic (in fact Bernoulli, Kolmogorov and ergodic), covering the unit 
circle uniformly, with
$\delta$-function correlation between consecutive $\theta$ values. Accordingly, we 
expect the growing cluster to be indistinguishable from a DLA, as is
indeed the case, cf. Fig.~\ref{Bercluster}. 

\begin{figure}
\hskip -1 cm
\epsfxsize=7.0truecm
\epsfysize=7.0truecm
\epsfbox{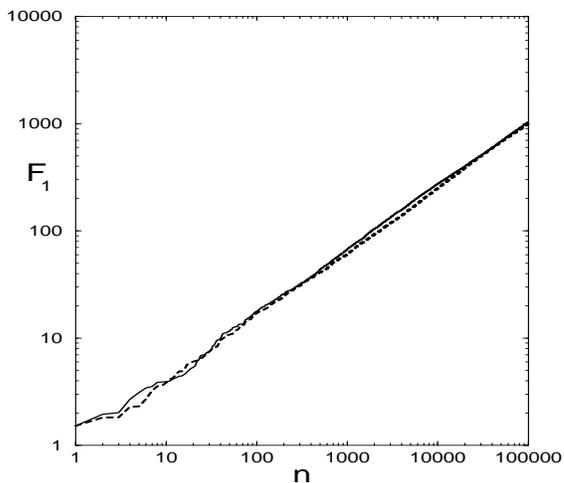}
\caption{Comparison of $F_1^{(n)}$ for a DLA (continuous line) and a Bernoulli itinerary
(dashed line).}
\label{F1DLABer}
\end{figure}

\begin{figure}
\hskip -1 cm
\epsfxsize=7.0truecm
\epsfysize=7.0truecm
\epsfbox{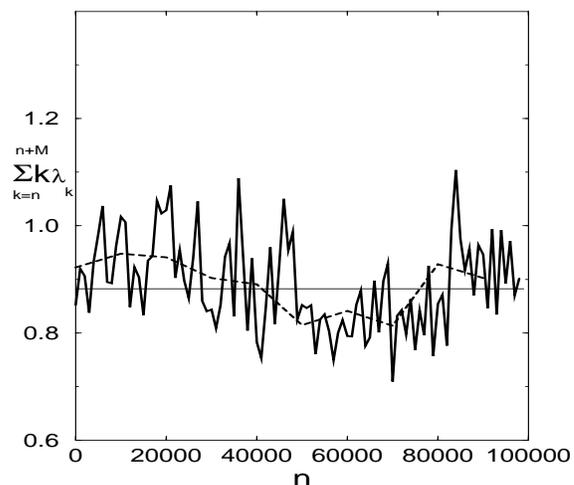}
\caption{The average of $k\lambda_k$ over the last $M$ iterations of the
Bernoulli itinerary, with $M=10^3$ (continuous line)  and $M=10^4$ 
(dashed line). 
The horizontal line is the expected value, $1/aD$. The fluctuations
are typical, reflecting the multi-scaling distributions of $\lambda_k$
in which large deviations are highly probable.}
\label{meanlamBer}
\end{figure}

One advantage of the present formalism
is that such a statement can be made quantitatively, not by eyeball.
The function $\Phi^{(n)}(w)$ and $\phi_{\lambda,\theta}(w)$ can be expanded in a
Laurent series in which the highest power is $w$ \cite{98HL,99DHOPSS}:

\begin{equation}
\Phi^{(n)}(w) = F^{(n)}_{1} w + F^{(n)}_{0} + F^{(n)}_{-1}w^{-1} +
F^{(n)}_{-2}w^{-2} + \dots 
\label{eq-laurent-f}
\end{equation}
The recursion equations for the Laurent coefficients of $\Phi^{(n)}(w)$
can be obtained analytically, and in particular one shows that \cite{98HL,99DHOPSS}
\begin{equation}
F_1^{(n)}  = \prod_{k=1}^{n}  [1+\lambda_k]^a  \,. \label{F1a}
\end{equation}
The importance of this lies in the fact that $F_1^{(n)}$ determines
that fractal dimension of the cluster. Defining $R_n$ as the minimal radius of 
all circles in $z$ that contain the $n$-cluster, one can prove that \cite{83Dur}
\begin{equation}
R_n\le 4F_1^{(n)} \ .
\end{equation}
Accordingly one expects that
\begin{equation}
F^{(n)}_1\sim n^{1/D}\sqrt{\lambda_0} \ ,
\label{eq-scalingrad}
\end{equation}
as $\sqrt{\lambda_0}$ is the only length scale in the problem.
We can thus present, as an example, plots of $F_1^{(n)}$
for our deterministic model (\ref{bernoulli}) together with $F_1^{(n)}$ in
any stochastic DLA growth, see Fig.~\ref{F1DLABer}. 
Another comparison is furnished by the statistics of $\lambda_n$. For the
DLA case it was shown in \cite{98HL,99DHOPSS} that 
\begin{equation}
\langle \lambda_n \rangle =\frac{1}{aDn} \ , \label{meanlam}
\end{equation}
where the average is taken over the harmonic measure. This is in agreement
with the ``electrostatic relation" derived by Halsey, \cite{87Hal}. In the Bernoulli 
itinerary there is no randomness and no probability measure, but
we may still define a ``running average" by, say, the last $M$ iterations
\begin{equation}
\langle \lambda_n \rangle_M \equiv \frac{1}{M}\sum_{k=n-M}^n \lambda_k \ . \label{meanM}
\end{equation}
In Fig.~\ref{meanlamBer} we show a related quantity, $(\sum_{k=n-M}^M k\lambda_k)/M$
for $M=1000$ and $M=10000$. We see that up to the expected fluctuations 
it settles down very
quickly on the appropriate value of the DLA cluster, i.e. $1/aD = .877..$.
Any other quantitative comparison that
one can think of leads to the same conclusion, i.e the Bernoulli itinerary
is a {\it bona fide} generic DLA. Of course, this is not surprising: the
correlation properties of successive values of $\theta_n$ in (\ref{bernoulli})
are indistinguishable from random numbers on the interval $[0,2\pi]$.
Nevertheless, our point is that the present growth algorithm gives us
freedom to choose deterministic itineraries resulting in DLA or other
growth patterns, and we next exploit this freedom to explore new geometries.

\section{Fractal Growth with quasi-periodic itineraries}
\subsection{Winding numbers and geometry}

\begin{figure}
\hskip -1 cm
\epsfxsize=7.0truecm
\epsfysize=8.0truecm
\epsfbox{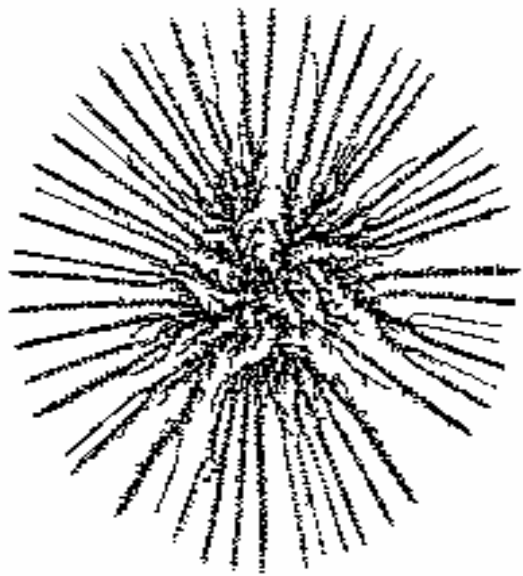}
\caption{Cluster grown with $W=233/144$ to $n=10^4$. Note the crossover from
fractal to 1-dimensional growth phases}
\label{233-144}
\end{figure}

\begin{figure}
\hskip -1 cm
\epsfxsize=7.0truecm
\epsfysize=8.0truecm
\epsfbox{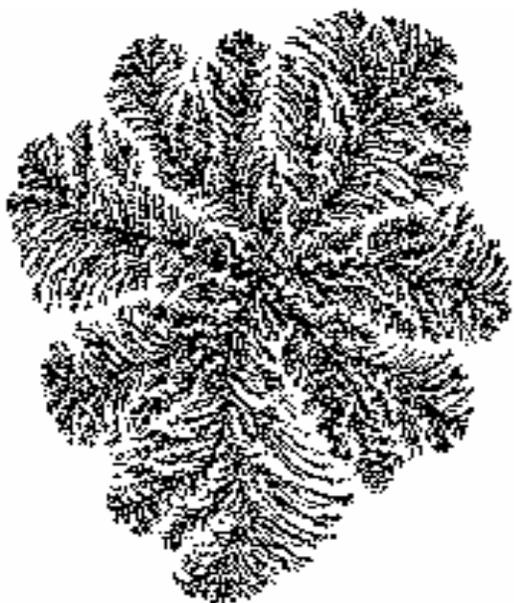}
\caption{The cluster grown with $W=\rho$ to $n=10^5$}
\label{gm100000}
\end{figure}

A new class of models is obtained by using a quasi-periodic itinerary.
Consider a simple map of the circle with a winding number W:
\begin{equation}
\theta_{i+1}=\theta_i+2\pi W \ . \label{winding}
\end{equation}
If we choose $W$ rational, $W=P/Q$, then after a cross-over time the cluster 
grown is locked into a
1-dimensional object made of rays. In the next subsection we present an
extensive discussion of the cross over time and of the properties of
the 1-dimensional phase of growth. As an example consider in
Fig.~\ref{233-144} the cluster resulting from (\ref{winding}) with $W=233/144$. 
On the other hand, for
an irrational winding $W$ the itinerary is ergodic and the cluster 
grown is geometrically non-trivial. As a first example we present the case
$W=\rho$ where $\rho$ is the Golden Mean, $\rho=(\sqrt{5}+1)/2$. The fractal
cluster that is associated with this rule is shown in Fig.\ref{gm100000}.
The cluster
has a fractal dimension $D= 1.86\pm 0.03$, as determined from the scaling
of $F_1^{(n)}$. This is considerably higher than DLA
(for which $D \approx 1.71$).

\begin{figure}
\hskip -1 cm
\epsfxsize=7.0truecm
\epsfysize=8.0truecm
\epsfbox{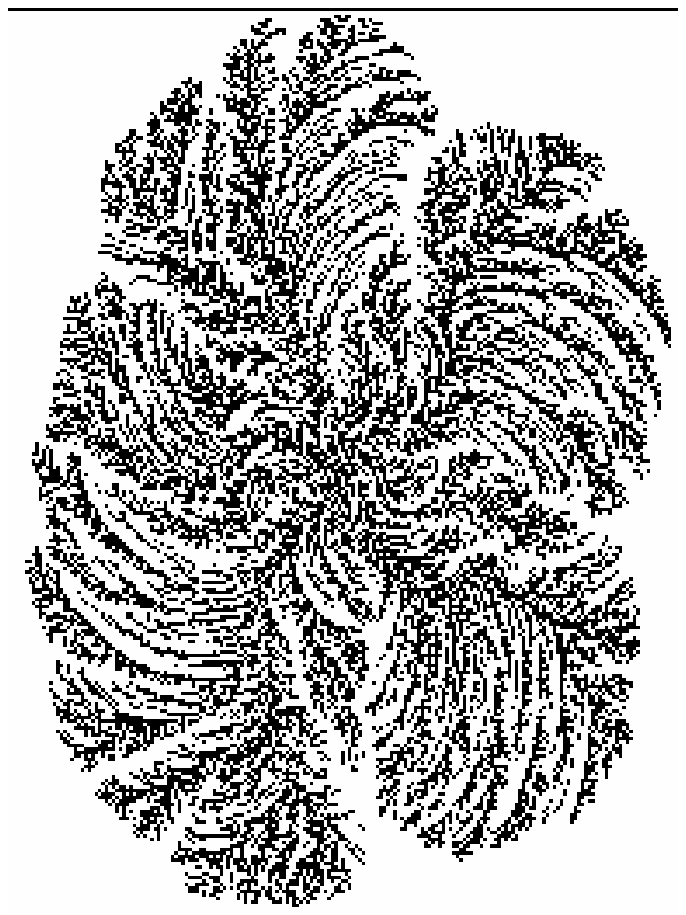}
\caption{The cluster grown with $W=2^{1/2}$ to $n=10^5$}
\label{sq2}
\end{figure}

\begin{figure}
\hskip -1 cm
\epsfxsize=7.0truecm
\epsfysize=8.0truecm
\epsfbox{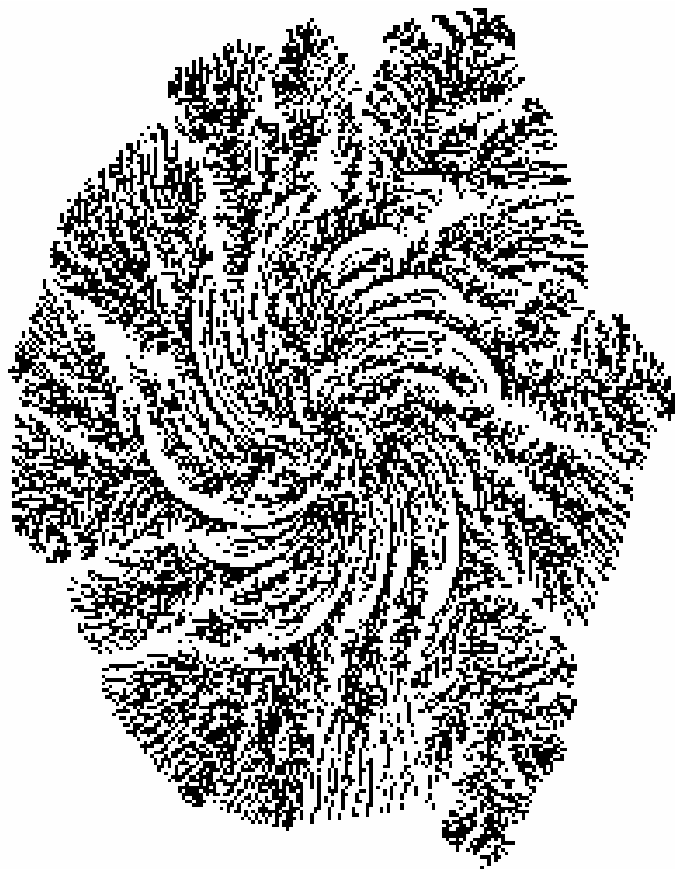}
\caption{The cluster grown with $W=3^{1/2}$ to $n=10^5$}
\label{sq3}
\end{figure}

\begin{figure}
\hskip -1 cm
\epsfxsize=7.0truecm
\epsfysize=8.0truecm
\epsfbox{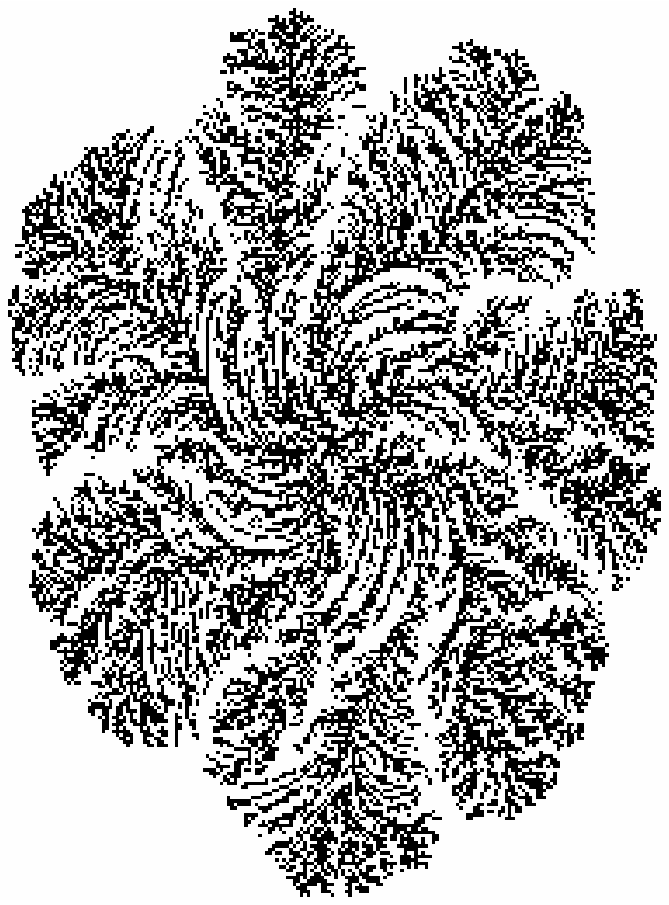}
\caption{The cluster grown with $W=(1+(10)^{1/2})/3$ to $n=10^5$}
\label{sq10}
\end{figure}

\begin{figure}
\hskip -1 cm
\epsfxsize=7.0truecm
\epsfysize=8.0truecm
\epsfbox{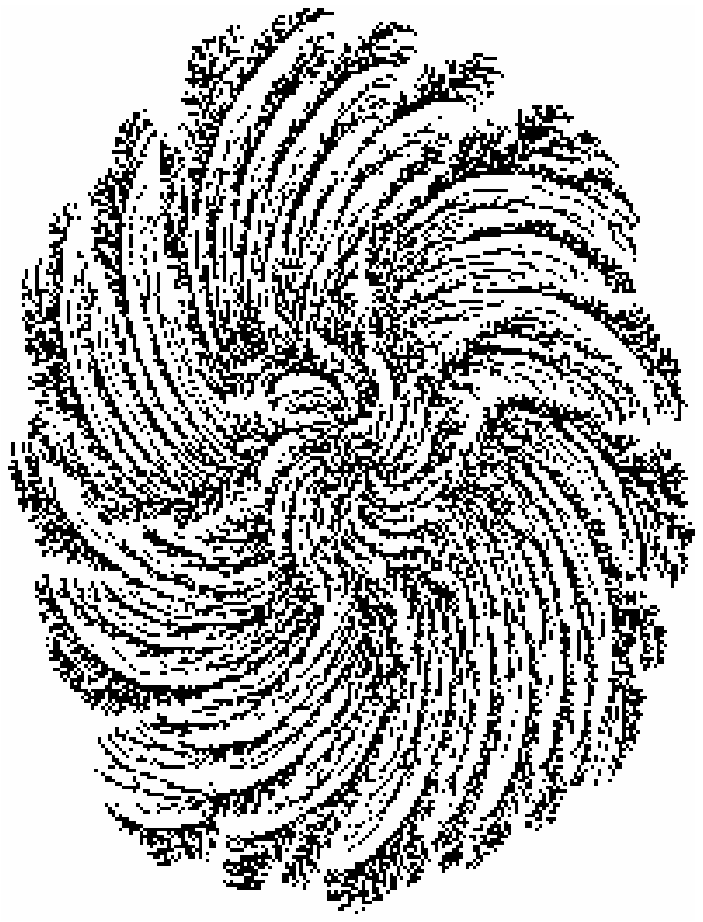}
\caption{The cluster grown with $W=((13)^{1/2}-1)/2$ to $n=10^5$}
\label{sq13}
\end{figure}

\begin{figure}
\hskip -1 cm
\epsfxsize=7.0truecm
\epsfysize=8.0truecm
\epsfbox{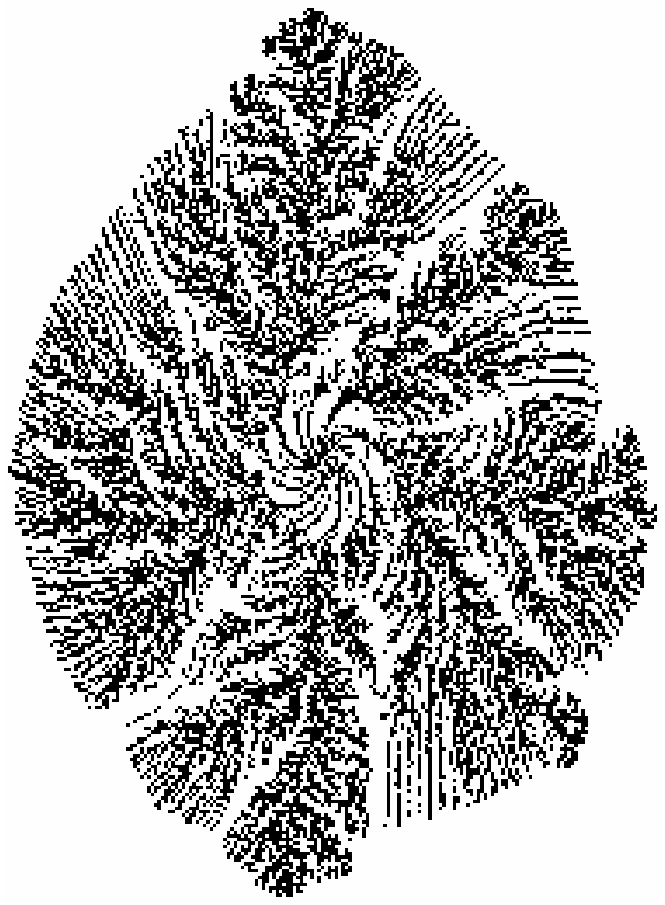}
\caption{The cluster grown with $W=7^{1/2}$ to $n=10^5$}
\label{sq7}
\end{figure}

The golden mean is best approximated by the continued fraction representation
\begin{equation}
\rho=\frac{1}{1+\frac{1}{1+ \frac{1}{1+\cdots}}} \ .
\end{equation}
Such a continued fraction is denoted below as $[ 0,\bar1]$. It is known
that the golden mean is special in presenting the slowest converging
continued fraction. Other quadratic irrationals also have periodic
continued fractions that converge faster. In Figs. \ref{sq2}-\ref{sq7} we show the
clusters grown with $W=\sqrt{2},~\sqrt{3},~(1+\sqrt{10})/3,~(\sqrt{13}-1)/2$
and $\sqrt{7}$ respectively. The continued fraction representations of these
winding numbers are $[1,\bar 2], [1,\overline{1,2}], [\overline{1,2,1}], [1,\bar 3], [2,\overline{1,1,1,4}]$
respectively. In choosing these examples we picked quadratic irrationals whose
representations converge relatively slowly. This facilitates the exposition of
scaling theory presented below.
%

We note that the clusters shown have very complicated geometry. Consider
for example the cases $W=(\sqrt{13}-1)/2$ and $W=\sqrt{7}$ shown in
Figs. \ref{sq13}, \ref{sq7} respectively. They exhibit thin spiral growth patterns
at their root, and then become bushy and thin in an apparently oscillatory
fashion. Accordingly, it becomes unclear whether the different quadratic
irrational winding numbers result in the same overall fractal dimension.
This question warrants some extra analysis. We will argue below that 
in spite of the difference appearance and the oscillations in the
``bushiness", the clusters grown by quadratic irrational winding
numbers have the same fractal dimension $D$.
\subsection{Different growth rules: period doubling itinerary}
\begin{figure}
\hskip -1 cm
\epsfxsize=7.0truecm
\epsfysize=8.0truecm
\epsfbox{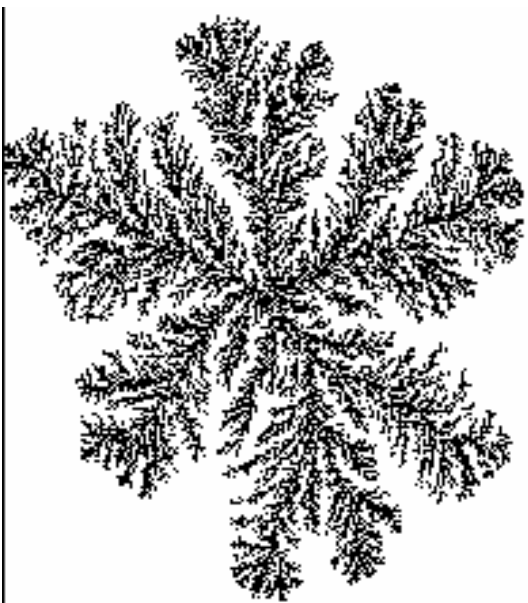}
\caption{The cluster grown with the period doubling itinerary to $n=10^5$}
\label{pdfig}
\end{figure}

\begin{figure}
\hskip -1 cm
\epsfxsize=7.0truecm
\epsfysize=8.0truecm
\epsfbox{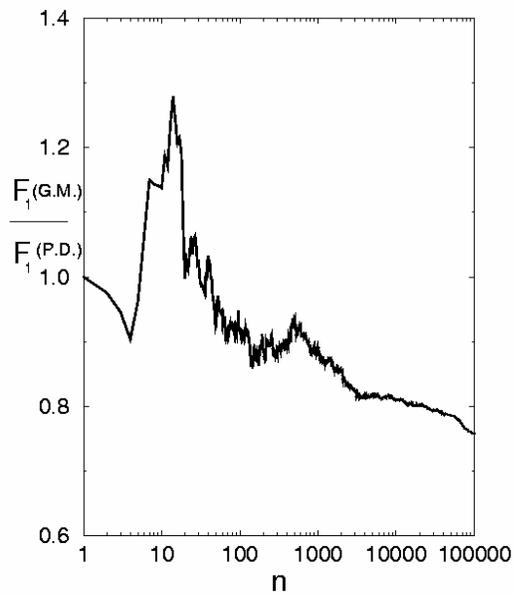}
\caption{The ratio between $F_1^{(n)}$ of the golden mean and period doubling
clusters.}
\label{compgmpd}
\end{figure}

Clearly, one can come up with an arbitrary number of different growth
rules. In this paper we will consider only one additional itinerary,
to underline the fact that quadratic irrational windings lead to 
a class of their own. This itinerary is constructed such that after every
$2^n$ iterations the points $\theta_k$ chosen on the circle
are equidistributed without repetitions. The order of visitation
is determined by the following rule:
\begin{eqnarray}
\theta_i&=&2\pi x_i\ , \nonumber\\
x_{i+1} &=& x_i+\frac{3}{2^{k_i+1}}-1\nonumber\\
k_i&=& -[ \log_2{(1-x_i)}] \ , \label{pd}
\end{eqnarray}
where $[\dots]$ stands for the integer value. We refer to this itinerary below
as the ``period doubling" algorithm.
The cluster grown with this rule is shown in Fig.~\ref{pdfig} . 
The dimension of this cluster is $D=1.77\pm 0.02$.
In contrast to the quadratic irrationals in this case a comparison
of $F_1^{(n)}$ of this cluster to $F_1^{(n)}$ of the golden mean itinerary 
shows a different scaling dependence on $n$ (cf. Fig.~\ref{compgmpd} ).

\subsection{Universality classes?}
In the previous section we noted that the geometry of some of the
clusters with quadratic irrational winding exhibit oscillations.
It is thus not clear whether they have the same fractal dimension
$D$. In this subsection we provide numerical test of the claim
that the quadratic irrationals belong to the same universality class.
In the the following sections we address this question using additional
tools.

\begin{figure}
\hskip -1 cm
\epsfxsize=7.0truecm
\epsfysize=8.0truecm
\epsfbox{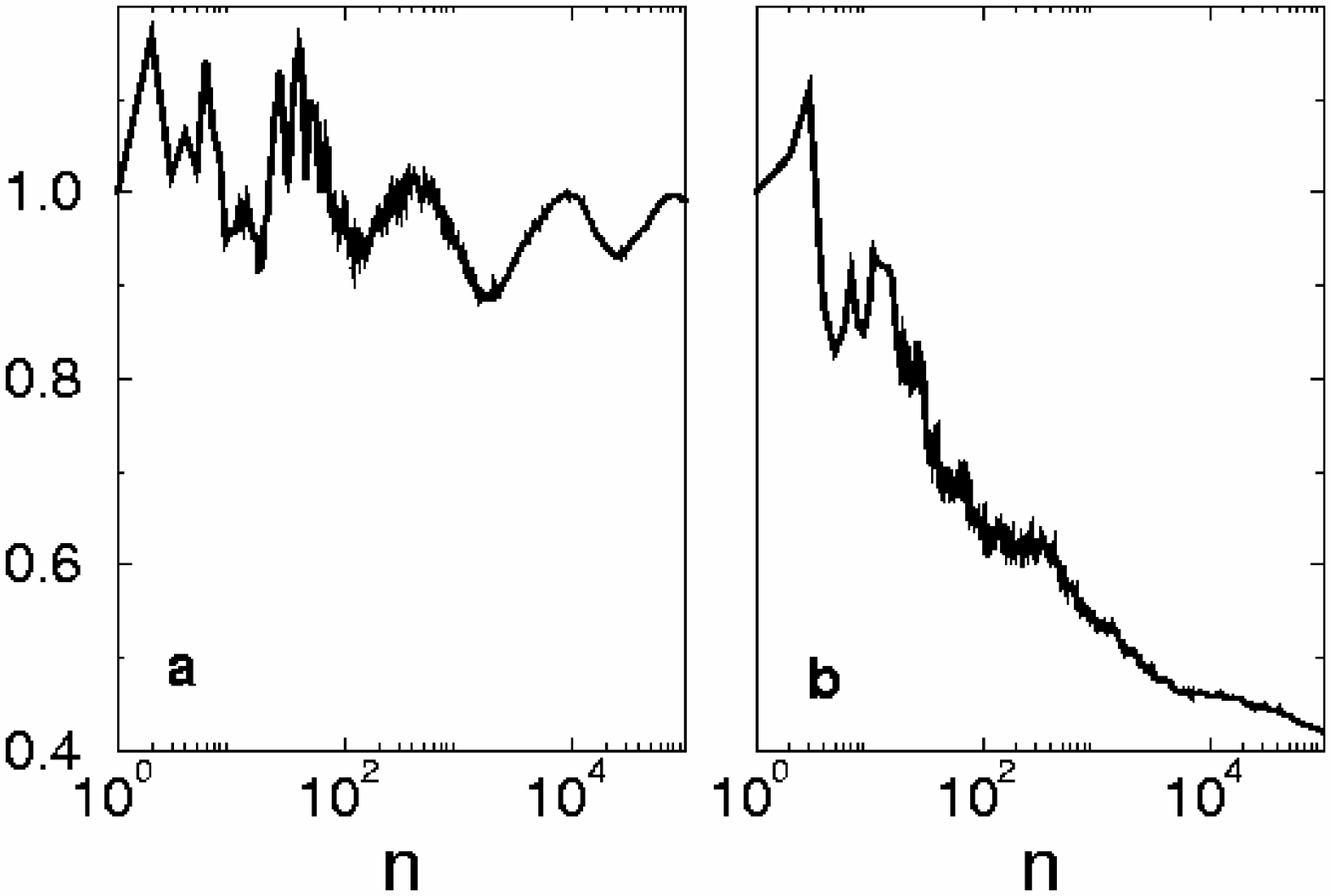}
\caption{Panel a: The ratio between of $F_1^{(n)}$ of the clusters grown 
with $W=2^{1/2}$ and
$W=3^{1/2}$. Panel b: The ratio between of $F_1^{(n)}$ of the cluster 
grown with $W=2^{1/2}$ and
a typical DLA}
\label{compgmsq2sq3}
\end{figure}

To study quantitatively
the oscillatory fractal geometry we consider the dependence of 
$F_1^{(n)}$ on $n$. In Fig.~\ref{compgmsq2sq3} panel a
we present compensated plots of $F_1^{(n)}(\sqrt{2})$ vs.$F_1^{(n)}(\sqrt{3})$ and vs. $F_1^{(n)}$
of a DLA as a function of $n$.

\begin{figure}
\hskip -1.5 cm
\vskip 1 cm
\epsfxsize=8.0truecm
\epsfysize=8.0truecm
\epsfbox{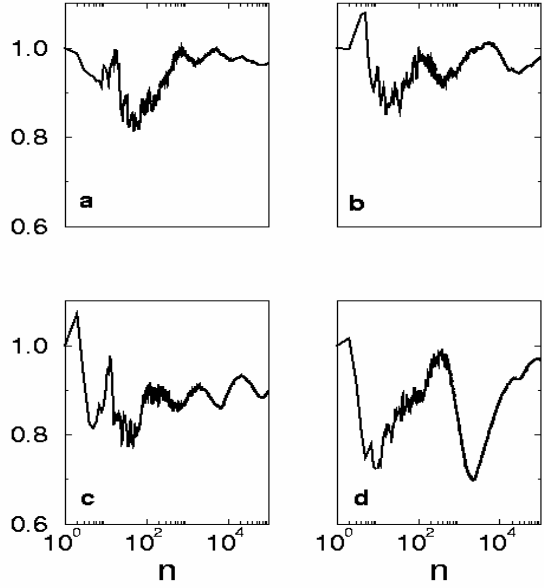}
\caption{The ratio between of $F_1^{(n)}$ of the golden mean and other 
quadratic
irrationals. Panels a-d show respectively $W=2^{1/2}$, $W=(1+(10)^{1/2})/3$
, $W((13^{1/2}-1)/2)$
and $W=7^{1/2}$}
\label{compgm-rest}
\end{figure}

It appears that although this ratio exhibits oscillations, these are bounded
and decreasing in amplitude, at least up to $n=10^5$. For comparison we show
in panel b of Fig.~\ref{compgmsq2sq3} a plot of $F_1^{(n)}(\sqrt{2})/F_1^{(n)}(DLA)$. Here
we see the clear difference in dimension as seen in the ratio approaching
zero as a power law in $n$. In Fig.~\ref{compgm-rest} we show
compensated plots of
$F_1^{(n)}$ of the clusters in Figs.~\ref{sq2}-\ref{sq7} 
versus $F_1^{(n)}$ of the golden mean growth. 
We see oscillations on the logarithmic scale, but again these are bounded, and
we propose that this points towards the possibility that all quadratic irrationals
winding numbers lead
to the same overall dimension of the cluster. In the next section we address
the issue of universality classes using additional tools. 
\section{Towards a scaling theory: winding with rational approximants}

To gain understanding of the geometry of the clusters grown with
quadratic irrational winding numbers we will make use now of the well
known fact that these irrationals can be systematically approximated 
by rational approximants. Thus, having a cluster constructed with a 
golden mean itinerary, a
natural question is what happens to the growth pattern when $\rho$ is replaced by 
ratios of
successive Fibonacci numbers  which are defined by the
recursion relation $F_{m+1}= F_{m}+F_{m-1}$, $F_0=0$, $F_1=1$. Using rational
approximants $\rho_m=F_{m-1}/F_m$, the itinerary becomes periodic on the unit circle
with period $F_m$, and it is observed in simulations (see Fig.~\ref{233-144} ) that
while for small clusters $n \ll n_c(F_m)$ the cluster appears fractal,
for $n \gg n_c(F_m)$ the cluster consists of a set of $F_m$ rays,
sometimes fused into a smaller set of 1-dimensional
rays whose number is extremely sensitive to the
initial conditions (here controlled by the value of $\lambda_0$).
\subsection{The 1-dimensional phase}

\begin{figure}
\hskip-1.5 cm
\vskip 2 cm
\epsfxsize=8.0truecm
\epsfbox{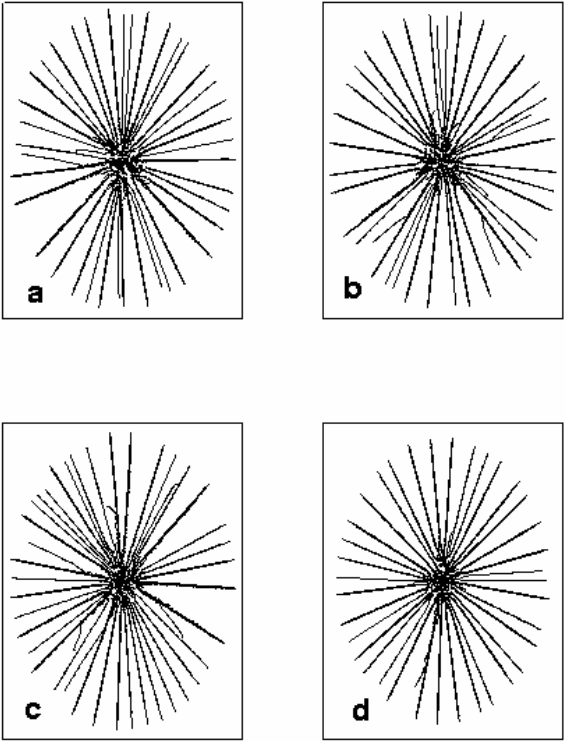}
\caption{Clusters grown with $W=144/89$ with four different values of
$\lambda_0$, from 0.11 to 0.88.}
\label{4lam0}
\end{figure}

\begin{figure}
\hskip-1 cm
\vskip 1 cm 
\epsfxsize=7.0truecm
\epsfbox{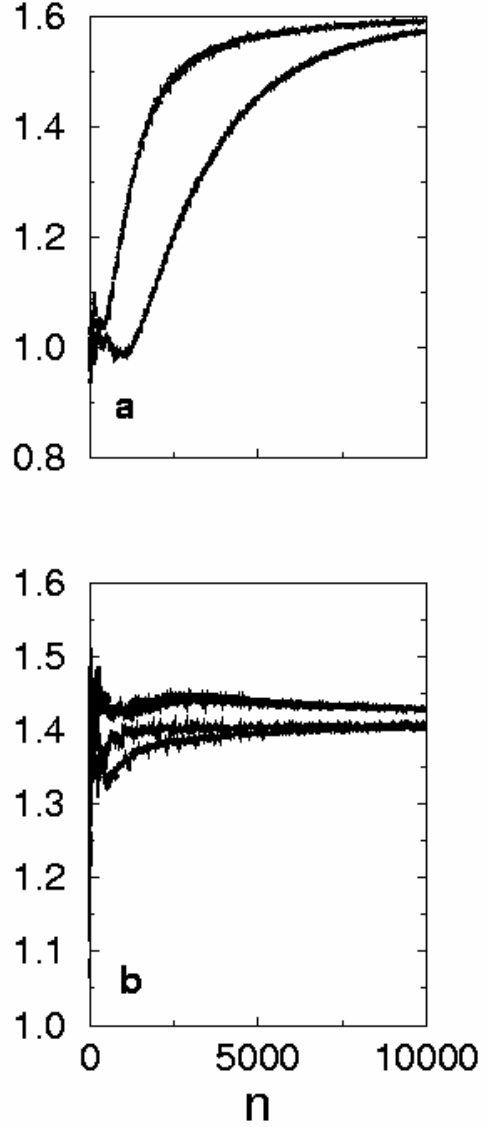}
\caption{Panel a: The ratio of $F_1^{(n)}$ for clusters grown with 
$\lambda_0=0.88$,
and winding numbers $W=89/55$ and $W=144/89$ (upper curve) and $W=144/89$ 
and $W=233/144$)
(lower curve). Both converge to the golden mean. Panel b: Similar plots with
$W=144/89$. $\lambda_0=0.22$ is compensated by $\lambda_0=0.11$, 0.44 by 
0.22 and 0.88 by 0.44.
All these plots converge to $2^{1/2}$.}
\label{lamF1}
\end{figure}

The properties of the 1-dimensional phase are important for developing
a scaling theory. As an example of the interesting behaviour seen as a
function of $\lambda_0$ consider Fig.~\ref{4lam0} in which clusters with $W=144/89$
are grown with 4 values of $\lambda_0$ which are 0.11, 0.22, 0.44, and 0.88. 
Evidently the cross over
from fractal to 1-dimensional behaviour depends on $\lambda_0$.
We also note
that the number of rays in the 1-dimensional phase has a nonmonotonic
dependence on $\lambda_0$. This indicates high sensitivity of the number
of rays to changes in the initial conditions. Obviously, the radius of the cluster
in the 1-dimensional case is inversely proportional to the number of rays.
On the other hand, we have found a surprising invariant: $F_1^{(n)}$ {\em is
asymptotically invariant to the number of rays (i.e. to initial conditions)
being always equal to $n\sqrt{\lambda_0}/Q$, up to a constant of proportionality
depending on the microscopic parameter $a$ only}. The numerical evidence is shown in
Fig.~\ref{lamF1}. Note the convergence to the golden mean in panel a, and 
to $\sqrt{2}$ in panel b (which is the value of the ratios of $\sqrt{\lambda_0}$).
This finding puts strict bounds on the number of possible rays.
The upper bound is obviously $Q$. The lower bound stems from the inequality
$R_n\le 4 F_1^{(n)}$, meaning that the number of rays must be larger than
$Q/4$. This invariance also indicates that the geometry of the rays is
not arbitrary, and that the angles between them are arranged to agree with
an invariant $F_1^{(n)}$.
\subsection{Scaling Function}

The crossover in fractal
shape is a general result for any periodic itinerary with $W=P/Q$, and
suggests the existence of a scaling for $F_1^{(n)}$ of the form

\begin{equation}
\label{scaling1}
F_1^{(n)} = n^{1/D} \sqrt{\lambda_0} f(n^{1/\alpha}/Q) ,
\end{equation}
where we have assumed that the crossover cluster size scales as
\begin{equation}
\label{crossover1}
n_c(Q) \sim Q^{\alpha}.
\end{equation}

The asymptotic forms of $F_1^{(n)}$ obey 
$F_1^{(n)}\sim n^{1/D} \sqrt{\lambda_0}$ for $n \ll n_c(Q)$, while
$F_1^{(n)} \sim (n/Q)\sqrt{\lambda_0}$ for $n \gg n_c(Q)$. In the first asymptote
we expect $D$ to be the same for all values of rational approximants to $\rho$,
including the limiting fractal cluster. The growing cluster cannot distinguish
between the rational approximant and the limiting irrational as long as
the fractal phase is observed. The second
asymptote is demonstrated in the previous subsection. Thus we require that
the asymptotic forms of the scaling function obey
\begin{eqnarray}
\label{asym1}
f(u) & \rightarrow & f(0)    \mbox{~as~$u \rightarrow 0$} \ , \\
f(u) & \sim     &    u     \mbox{~as~$u \rightarrow \infty$} \ . \label{asym2}
\end{eqnarray}
The second asymptote (\ref{asym2}) determines the scaling relation
\begin{equation}
\alpha = D/(D-1) \ .\label{alfa}
\end{equation}
For the Golden Mean fractal $D
\approx 1.86$ and consequently in this case $\alpha \approx 2.16$.

\begin{figure}
\hskip -1 cm
\epsfxsize=7.0truecm
\epsfysize=8.0truecm
\epsfbox{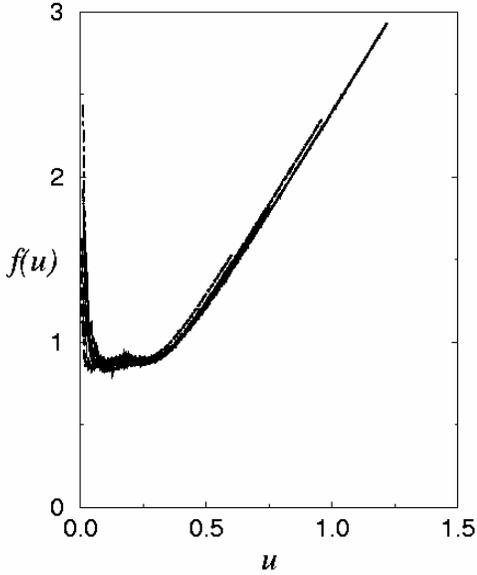}
\caption{Scaling behaviour for six separate data sets for $f(u)$ with $u = n^{.46}/Q$.
Shown are $\lambda0=0.88$ with $W=89/55$, $W=144/89$, $W=233/144$, $W=377/233$,
$\lambda0=0.44$ with $W=144/89$, $\lambda0=0.22$ with $W=144/89$.}
\label{collapse}
\end{figure}

In Fig.~\ref{collapse}  $F_1^{(n)}/(n^{1/D} \sqrt{\lambda_0})$ is plotted
against the scaling variable $u = n^{1/\alpha}/Q$ for
six different clusters with different values of $W$ and $\lambda_0$. 
The best data
collapse was obtained using the value $\alpha = 2.15$
The data collapse achieved is readily apparent with the scaling function 
$f(u)$ predicted by the theory.
\section{The crossover and the estimate of the dimension}

In this section we discuss the properties of the conformal map $\phi_{\lambda,\theta}$
which determine the cross over from fractal to 1-dimensional growth. In other words,
we will attempt to provide an independent estimate of $n_c$ as a function
of the winding number $W$. If we succeeded to estimate the exponent $\alpha$ in
(\ref{crossover1}) independently from Eq.(\ref{alfa}), we would have an equation
for the dimension.

To understand the crossover, we note that the reason for the fractal growth
phase with rational winding is that after every event of growth the interface
$z(e^{i\theta})$ is non-locally reparametrized in addition to the local
growth event. Accordingly, a periodic orbit on the unit circle is not necessarily
mapped to a periodic orbit in $z$. The region in the unit circle which is significantly
affected by growing the $n$th bump has a scale $\sqrt{\lambda_n}$ centered around
$\theta_n$ \cite{94Hal}. Accordingly
we can estimate when reparametrization will cause a ``miss" in the mapped
orbit: as long as
\begin{equation}
\sqrt{\lambda_n}\ge \frac{2\pi}{Q} \ , \label{condition}
\end{equation}
the growth will remain fractal. We can therefore expect a crossover to
1-dimensional growth when this condition is violated, something that
is bound to happen since typical values of $\lambda_n$ are expected to
decreases with $n$, cf. Eq.(\ref{meanlam}) and the discussion below.

What remains is to estimate $\lambda_n$ as a function of $n$ in the cross
over region that is defined by 
\begin{equation}
\sqrt{\lambda_{n_c}}\approx 2\pi/Q \ . \label{lamnc}
\end{equation}
 In the fractal
region $\lambda_n$ is a highly erratic function of $n$. Even though we
do not have here randomness in the sense of DLA, it is natural to consider,
in a fashion similar to Eq.(\ref{meanM}),
the distribution of $\lambda_k$ over $Q$ successive steps of growth. For $Q$ 
large enough such distributions have well defined moments. In particular
consider the first moment
\begin{equation}
\langle \lambda_n \rangle_Q \equiv \frac{1}{Q} \sum_{k=n-Q}^n \lambda_k
\end{equation}
The power law dependence of $F_1^{(n)}$ and Eq.(\ref{F1a}) imply that this
moment has to be
\begin{equation}
\langle \lambda_n \rangle_Q =\frac{1}{a n D} \ . \label{lamnmean}
\end{equation}

\begin{figure}
\hskip -1 cm
\epsfxsize=7.0truecm
\epsfysize=8.0truecm
\epsfbox{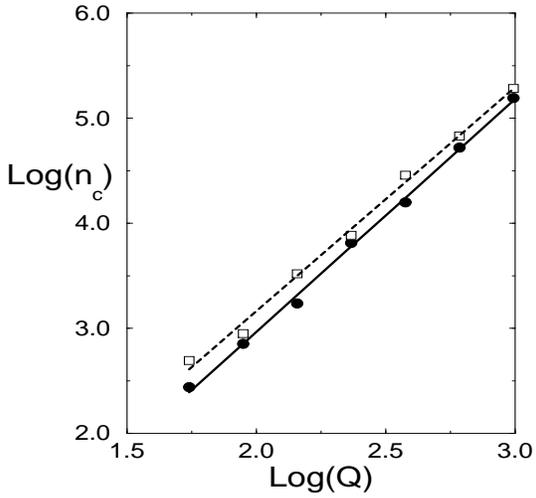}
\caption{The measured the cross over values $n_c$ as a function of $Q$
in a log-log plots. In dots are the values of $n$ for which the condition 
$\lambda_n^{\rm max}=4\pi^2/Q^2$ was met for the first time, in squares for 
the last time.}
\label{nc}
\end{figure}

If we estimate $\lambda_{n_c}$ in Eq. (\ref{lamnc}) by its mean (\ref{lamnmean}), 
we would write
\begin{equation}
\lambda_{n_c} \sim 1/n_c ~~\rightarrow~~ n_C\sim Q^2.
\end{equation}
Thus $D/(D-1)=2$ or $D=2$. Even though we get an overestimate, this is
a good indication that we are on the right track. The reason for the overestimate
is that we neglected the fluctuations that sometime lead to $\lambda_n$ much
larger than the mean. We expect a cross over to occur when the {\em largest}  
$\sqrt{\lambda_k}$ are smaller than $2\pi/Q^2$, since it is enough to have a few
large $\lambda_k$ to cause a reparametrization that will ruin a potential periodic orbit.
 We thus seek a condition
\begin{equation}
\lambda_n^{\rm max}\equiv {\rm max} \{\lambda_k\}_{k=n-Q}^n \approx \frac{4\pi^2}{Q^2} \ . \label{zehu}
\end{equation} 
We note that $\lambda_k$ is an erratic function of $k$, and therefore the condition
(\ref{zehu}) can be met more than once in a given series $\lambda_k$. In Fig.~\ref{nc}  we
show two log-log plots of
$n_c$ computed from the value of
$n$ for which $\sqrt{\lambda_n^{\rm max}}=2\pi/Q$, plotted as a function of $Q=F_m$. 
The cross over value $n_c$
was computed in two different ways. In circles we exhibit the values obtained from
measuring when $\sqrt{\lambda_n^{\rm max}}=2\pi/Q$ {\em for the first time}, and
in squares we exhibit the values obtained from $\sqrt{\lambda_n^{\rm max}}=2\pi/Q$ {\em for the last
time}. Computing the slopes by linear regression and averaging 
between them we find the scaling law
\begin{equation}
n_c\sim Q^{2.17\pm0.03} \ .
\end{equation}
Comparing with Eqs.(\ref{crossover1}), (\ref{alfa}) we get an estimate for $D=1.86\pm 0.03$,
in excellent agreement with the determination of the dimension by $F_1^{(n)}$.

We note in passing that $\lambda_n^{\rm max}$ can be assigned a generalized
dimension $D_\infty$ in the language of Hentschel and Procaccia \cite{83HP}. 
Define
\begin{equation}
\langle \lambda_n^q \rangle_Q \equiv \frac{1}{Q} \sum_{k=n-Q}^n \lambda^q_k \ .
\end{equation}
From \cite{99DHOPSS} the precise scaling law is
\begin{equation}
\lambda_n^{\rm max} = \lim_{q\to \infty} \langle \lambda_n^q \rangle_Q ^{1/q}  \sim
n^{-2D_\infty/D}
\ .
\end{equation} 
Comparing with (\ref{alfa}) we conclude that in this case there exists
a scaling relation
\begin{equation}
D_\infty =D-1 \ .
\end{equation}
Such a scaling relation was conjectured by Turkevich and Scher for DLA \cite{85TS}
(of course with a different $D$ and $D_\infty$). While
there are severe doubts about the correctness of this conjecture
for DLA\cite{87Hal}, we point out that in our case it follows directly from
elementary considerations.
\subsection{The period doubling itinerary}

\begin{figure}
\hskip -1 cm
\epsfxsize=7.0truecm
\epsfysize=8.0truecm
\epsfbox{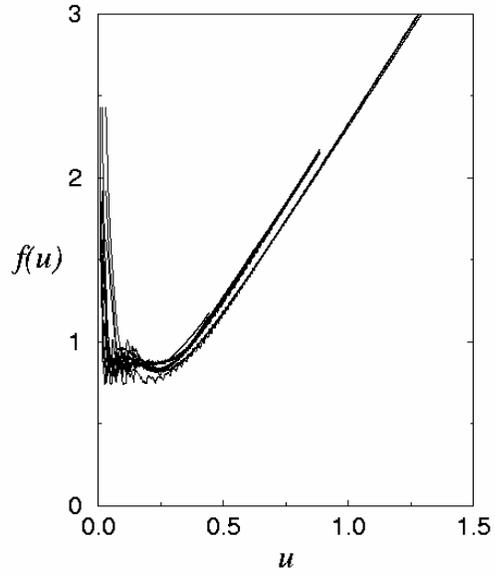}
\caption{Scaling behaviour for nine separate data sets for clusters 
grown with truncated period doubling itineraries, in addition to one 
data set of the quasiperiodic class. See text for details}
\label{collapsepd}
\end{figure}

Even though the period doubling itinerary leads to a cluster
whose fractal dimension differs from the quadratic irrational windings,
we show here that the ideas presented above pertain equally to this 
growth pattern. Instead of rational approximants we use here, naturally,
$2^n$-periodic orbits which are obtained by cutting the itinerary (\ref{pd})
after $2^n$ iterations and repeating it periodically. The crossover
from fractal to 1-dimensional growth is seen also in this case, and
we can use it in a very similar way to identify the crucial exponent that
determines the dimension of the asymptotic cluster.
Indeed, the whole set of ideas developed above repeats verbatim by
changing $Q$ with $2^n$. What remains is to find $\lambda_n^{\rm max}$
as a function of $n$. 
In Fig~.\ref{collapsepd} we show the data collapse 
obtained as in Fig.~\ref{collapse}
for the quasiperiodic analog. We show nine different data sets with 
periodic itineraries of periods 32, 64 and 128 and $\sqrt{\lambda_0}$ values
of 0.22, 0.44 and 0.88. The scaling function for these data sets is
plotted as a function of $u = n^{.44}/Q$, where the exponent is computed 
from $D=1.78$. It is noteworthy that the scaling function obtained 
appears identical to the scaling function $f(u)$ for the quasiperiodic 
family. For comparison we
added in Fig~.\ref{collapsepd} also one curve from the quasiperiodic
class, and it appears indistinguishable from the rest.  

\begin{figure}
\hskip -1 cm
\epsfxsize=7.0truecm
\epsfysize=8.0truecm
\epsfbox{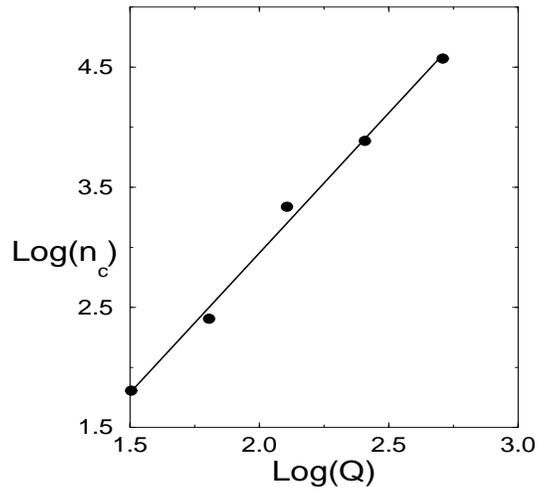}
\caption{The measured the cross over values $n_c$ as a function of $Q=2^n$
in a log-log plot. Shown are the values of $n$ for which the condition 
$\lambda_n^{\rm max}=(2\pi/2^n)^2$ was met for the first time.}
\label{ncpd}
\end{figure}

The conclusion from this data collapse is that the mechanism governing
the crossover from fractal to 1-dimensional growth phases here is the
same as the one discussed above for the quasiperiodic itineraries.
The difference between the dimensions of
the period doubling cluster and the quasiperiodic cluster must lie
in the different numerical value of the exponent characterizing
$\lambda_n^{\rm max}$ as a function of $n$. In this case the natural
averaging cycles are of length $Q=2^n$. Fig.~\ref{ncpd} is the 
analog of Fig.~\ref{nc} for the period doubling
itinerary, where the critical value $n_c$ was estimated from the
first time that $\sqrt{\lambda_n^{\rm max}}$ became smaller than $2\pi/2^n$.

The linear regression provides us with the the
scaling law
\begin{equation}
n_c\sim Q^{2.33\pm 0.1} \ .
\end{equation}
Computing $D$ we find $D=1.75\pm 0.05$ in good agreement with the numerical
estimate from $F_1^{(n)}$. 
\section{Summary and the road ahead}

The main points of this paper are as follows:
\begin{itemize}
\item The iterated conformal maps algorithm for fractal growth patterns
offers a convenient way to introduce a large number of deterministic growth models
with highly non-trivial fractal geometry.
\item Itineraries with irrational winding numbers generate fractal growth
patterns. We proposed that all the quadratic irrationals produce clusters of
the same fractal dimension, in spite of different appearance.
\item By considering a series of rational approximants we could produce a
scaling theory of the growing clusters, achieving data collapse for all
values of $n,\lambda_0$ and $P/Q$. 
\item Identifying the mechanism for the cross over from fractal to 1-dimensional
growth phases we could pinpoint the exponent that determines the fractal dimension
$D$. This exponent characterizes the $n$ dependence of the extremal values
of $\lambda_n$. 
\item The mechanism appears general; itineraries leading to different 
cluster dimensions, like the period doubling itinerary (\ref{pd})
and its truncated versions, can be understood in the same way. The
scaling function (\ref{scaling1}) and the scaling relation (\ref{alfa})
are general, but the exponent $\alpha$ changes. Its determination
by the scaling of $\lambda_n^{\rm max}$ Eq.(\ref{zehu}) is however general.
\end{itemize}
We note that all the numerical tests point out in favour of this scenario,
and in our opinion rule out a value $D=2$ for the clusters discussed above.
The only way to get 2-dimensional growth, as shown
above, is if the distribution of $\lambda_n$ does not multiscale, i.e. 
all $D_q$ are the same, and the scaling of $\lambda_n^{\rm max}$ identifies with the
scaling of the average of $\lambda_n$.

Nevertheless, we point out that the crucial step in our scenario,
the determination of the exponent $\alpha$ in Eq.(\ref{crossover1}),
was achieved numerically. The scaling theory presented above has a strong
flavour of a renormalization group approach. It appears that such
an underlying theory may have a low codimension, maybe with 1
important exponent, the one characterizing the rate of crossover
of the rational approximants to the irrational limit. The search
of such a theory appears to be an important task for the near
future.

\acknowledgments
We benefitted from discussions with T.C. Halsey, 
C. Tresser and L. Peliti.
This work has been supported in part by the
European Commission under the TMR program and the Naftali and Anna 
Backenroth-Bronicki Fund for Research in Chaos and Complexity.




\newpage

\narrowtext
\narrowtext
\narrowtext
\narrowtext
\narrowtext
\narrowtext
\narrowtext
\narrowtext
\narrowtext
\narrowtext
\narrowtext
\narrowtext
\narrowtext
\narrowtext
\narrowtext
\narrowtext
\narrowtext
\narrowtext
\narrowtext
\narrowtext

\end{document}